# A review on the complementarity of renewable energy sources: concept, metrics, application and future research directions

by: J. Jurasz[1,2,*], F.A. Canales[3], A. Kies[4], and M. Guezgouz[5], A. Beluco[6]

[1] School of Business, Society and Engineering, Future Energy Center, Mälardalen University, 72123 Västerås, Sweden

[2] Faculty of Management, Department of Engineering Management, AGH University, 30 Mickiewicza Ave., 30-059 Cracow, Poland

[3] Department of Civil and Environmental, Universidad de la Costa, Calle 58 #55-66, 080002 Barranquilla, Atlántico, Colombia

[4] Frankfurt Institute for Advanced Studies, Goethe University Frankfurt, 60438 Frankfurt am Main, Germany

[5] Department of Electrical Engineering, Mostaganem University, BP188/227, Mostaganem 27000, Algeria

[6] Instituto de Pesquisas Hidráulicas, Universidade Federal do Rio Grande do Sul, Av Bento Goncalves, 9500, Caixa Postal 15029, Bairro Agronomia, 91570-901 Porto Alegre, Rio Grande do Sul, Brazil

*jakub.jurasz@mdh.se

**Abstract**: It is expected, and regionally observed, that energy demand will soon be covered by a widespread deployment of renewable energy sources. However, the weather and climate driven energy sources are characterized by a significant spatial and temporal variability. One of the commonly mentioned solutions to overcome the mismatch between demand and supply provided by renewable generation is a hybridization of two or more energy sources in a single power station (like wind-solar, solar-hydro or solar-wind-hydro). The operation of hybrid energy sources is based on the complementary nature of renewable sources. Considering the growing importance of such systems and increasing number of research activities in this area this paper presents a comprehensive review of studies which investigated, analyzed, quantified and utilized the effect of temporal, spatial and spatio-temporal complementarity between renewable energy sources. The review starts with a brief overview of available research papers, formulates detailed definition of major concepts, summarizes current research directions and ends with prospective future research activities. The review provides a chronological and spatial information with regard to the studies on the complementarity concept.

Highlights

- Pearson correlation is most common metric of complementarity quantification
- Concept of complementarity is often mentioned but clear application is not provided
- Most research activities focus on: Brazil, Europe, China and USA

# 1. Introduction and motivation

Over the last years, variable renewable energy sources (VRES) have become a cost competitive and environment-friendly alternative to supply power to isolated and large-scale power grids around the globe. Nevertheless, because of their intermittent/variable/stochastic/non-dispatchable characteristic, they cannot provide the grid with various additional and mandatory services other that delivering a certain volume of energy (however, it is worth mentioning that besides active power, inverters of renewable power generators can also provide reactive power for voltage control and could be programmed to provide inertia [1]). In order to explore how to effectively improve VRES integration into the power systems, more needs to be known about the underlying behavior patterns and dynamics of their power generation. Over the recent years many investigations have been focus on this VRES grid integration [2], [3], [4], [5], [6], [7].

Several solutions worth mentioning were presented by Jacobson and Delucchi (2011) [8] who suggested to apply them in order to facilitate the process of VRES integration:

- interconnecting spatially distributed generators;
- using complementary or/and dispatchable generators in hybrid configurations;
- application of demand-response and flexible loads;
- deploying energy storage;
- oversizing and power to hydrogen;
- using the concept of vehicle to grid – use of electric vehicles as storage;
- forecasting of VRES generation.

Notably, two of the referenced concepts mention the use of a combination of VRES sources, which exhibit a complementary nature of their operation. First is the hybridization of energy sources (like solar-wind, wind-hydro etc.) and the second is the use of spatial distribution of generators to smooth the power output of given VRES. Both concepts are based on the complementary (to various extent) nature of renewable energy sources.

From the literature point of view, we would like to draw attention to works by Hart et al., (2012) [9] and Engeland et al., (2017) [10]. Although not explicitly, both considered the concept of VRES complementarity in their analysis. The first paper provided an in-depth analysis from the power system operation perspective, whereas the second looked at it more from a climate/meteorological perspective.

The objective of this review paper is to: provide a thorough overview on the past research activities concerning the concept of energetic resources complementarity; analyze and assess the existing state of knowledge; provide guidelines for potential future research directions. More specifically, we aim at answering the following research questions:

- What is and how is the concept of energetic complementarity defined?
- Which are the types of energetic complementarity?
- How did the studies on complementarity evolve and what are major, still unresolved shortcomings?
- Which metrics/indices are used to evaluate complementarity?
- What are potential applications of complementarity metrics?

In our research, we have applied a narrative approach of writing a review paper. Consequently, we have focused on comparing, and summarizing the existing theory/models and formulating conclusions on qualitative level. The investigated topic has been structured based on methodological approaches (indices and methods for complementarity assessment), chronological order and geographical location. To ensure a wide coverage of studies investigating complementarity concept we have used the following search engines: Science Direct, Scopus and Google Scholar. After the initial screening of papers containing the keyword "complementarity", we have performed an additional search based on the references present in those papers and included those relevant to the subject.

According to the Cooper's taxonomy [11] this review can be described as summarized in Table 1.

*Table 1. Characteristics of this review based on Cooper's taxonomy*

| Characteristic | Cooper's definition | Authors' selection |
|---|---|---|
| Focus | The material that is of central interest of the reviewer | Research methods, outcomes, theories and applications |
| Goal | What the author hopes the review will reveal | Integration of available knowledge, criticism |
| Perspective | Reviewer point of view | Neutral |
| Coverage | The extent to which reviewer includes the relevant works | Representative |
| Organization | Paper organization | Methodological |
| Audience | Intended paper audience | Specialized scholars/decision makers |

The reminder of this paper is structured as follows: in Section 2 we start with a clear and updated definition of the "complementarity" concept. In Section 3 we present the historical and geographical overview of the research on the complementarity – simply statistics on complementarity research. In Section 4 we analyze and describe the various metrics used to assess the complementarity. In Section 5 we discuss the data source used by the authors and current state of the research. In Section 6, we discuss current possibilities of applying the concept of complementary and we formulate further potential applications. The paper ends with Section 7, which summarizes the review and presents potential future research directions.

## 2. Definition of the complementarity concept

Based on the literature review, the following definition can be built up around the term complementarity. According to the Oxford dictionary, the term complementarity is: "a relationship or situation in which two or more different things improve or emphasize each other's qualities". Considering the context of energy sources, the complementarity should then be understood as the capability of working in a complementary way. Complementarity can be observed in time, space and jointly in both domains. Besides the graphical explanations in Figure 1 and Figure 2, the following paragraphs provide their brief definition:

**Spatial complementarity** – can be observed between one or more types of energy sources. It is a situation when energy resources complement each other over certain region. Scarcity of one VRES in region *x* is complemented by its availability in region *y*. An example of space complementarity can be the smoothing effect of spatially distributed wind generators whose energy production trends exhibit decreasing coefficient of correlation with an increasing distance between sites.

**Temporal complementarity** – can be observed between two or more energy sources in the same region. It is understood as a phenomenon when VRES exhibit periods of availability which are complementary in the time domain. As an example, it is possible to mention the annual patterns of wind and solar energy availability over Europe, where the former is abundant in Autumn – Winter whereas the latteris abundant in the Spring-Summer period. An example of temporal complementarity for a single source is provided in the note below Table 2. This table presents the characteristics of the different types of complementarity.

**Spatio-temporal complementarity** - (complementarity in time and space) is considered for a single or multiple energy sources whose complementary nature is investigated simultaneously in time and space domains. A good example is the Brazilian power system and its hydropower resources, which lead to an interconnection of the south-southeastern and north-northeastern subsystems.

The energetic complementarity can be assessed based on various indices and metrics, with the most relevant being described in section 5.

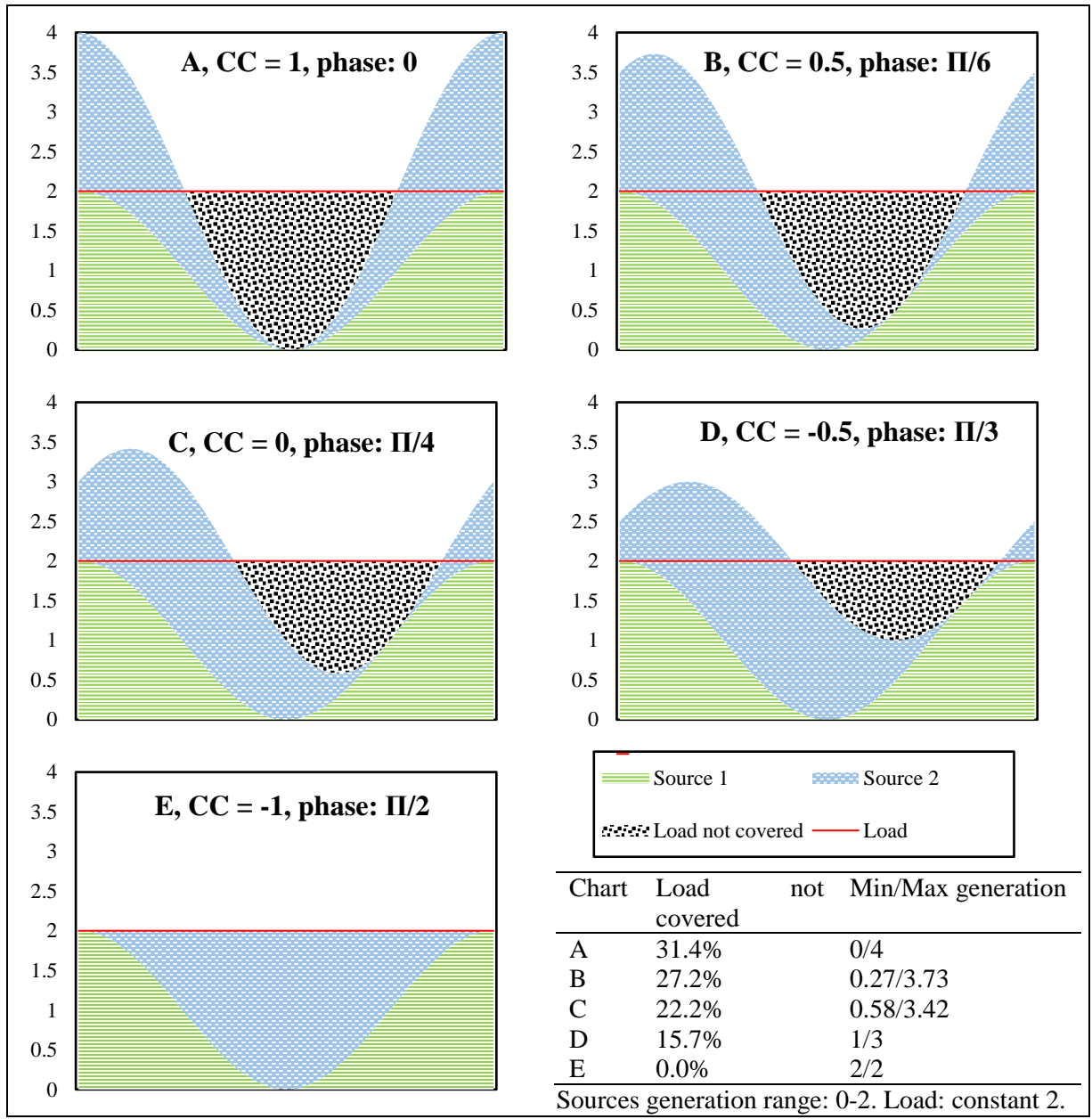

*Figure 1. Complementarity concept explained by means of sine signal. CC – coefficient of correlation.*

*Table 2. Characteristics of the different types of energetic complementarity*

| Type of complementarity | Number of sources considered | Number of sites/regions considered | Factor driving the existence of complementarity |
|---|---|---|---|
| Temporal | ≥2* | =1 | Different availability in time |
| Spatial | ≥1 | ≥2 | Different availability in space |
| Spatio/temporal | ≥1 | ≥2 | Time/space different availability |

*In case of temporal complementarity, a single energy source can be also considered by using the "flexibility" offered by technology. For example, the complementarity (smoother power output over the day/year) of single PV system can be increased by mounting PV arrays at different azimuths and inclination angles. The same applies to the wind farm where different wind turbines can be used with various hub heights or power curves.

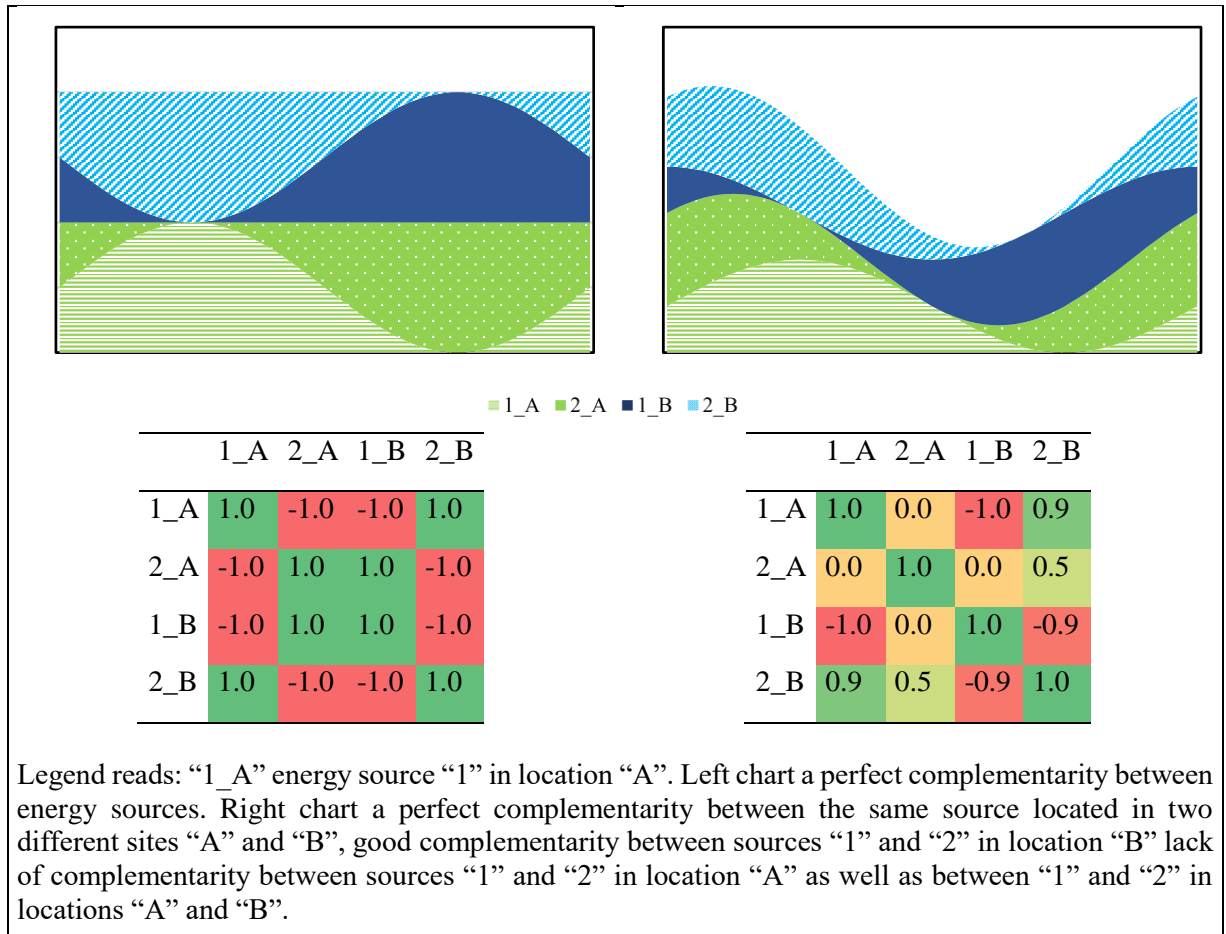

Legend reads: "1_A" energy source "1" in location "A". Left chart a perfect complementarity between energy sources. Right chart a perfect complementarity between the same source located in two different sites "A" and "B", good complementarity between sources "1" and "2" in location "B" lack of complementarity between sources "1" and "2" in location "A" as well as between "1" and "2" in locations "A" and "B".

*Figure 2. Conceptual visualization of spatio/temporal complementarity.*

## 3. Historical and research spatial distribution overview

We have performed an in-depth analysis of available literature on the complementarity concept and found that the first papers dedicated to this topic were published in the late-seventies. Two papers by Kahn [12], [13] and a paper by Takle and Shaw [14] investigated complementarity, mostly focusing on reliability of spatially distributed wind generators. The aim of their works was to answer the question whether spatial dispersed wind parks can provide firm power to the system, thus allowing replacing conventional fuel generation. Kahn stated that the study of energetic complementarity is a very promising area of research, and its assessment should be considered when pondering VRES integration to power systems. On the other hand, the paper by Takle and Shaw investigated the temporal complementarity between solar and wind resources in Iowa. Their findings show strong complementary on an annual basis but only slight one a daily scale. Along with the series of studies summarized by Justus and Mikhail (1979) [15], the aforementioned works can be considered as the initial research contributions in the area of the complementarity between renewable sources. Over the recent years the number of papers dealing with the complementarity concept widely varied, however, the period from 2016 to 2018 has been the most productive, as seen in Figure 3. This might be associated with the growing importance of renewable sources and their integration in the power systems.

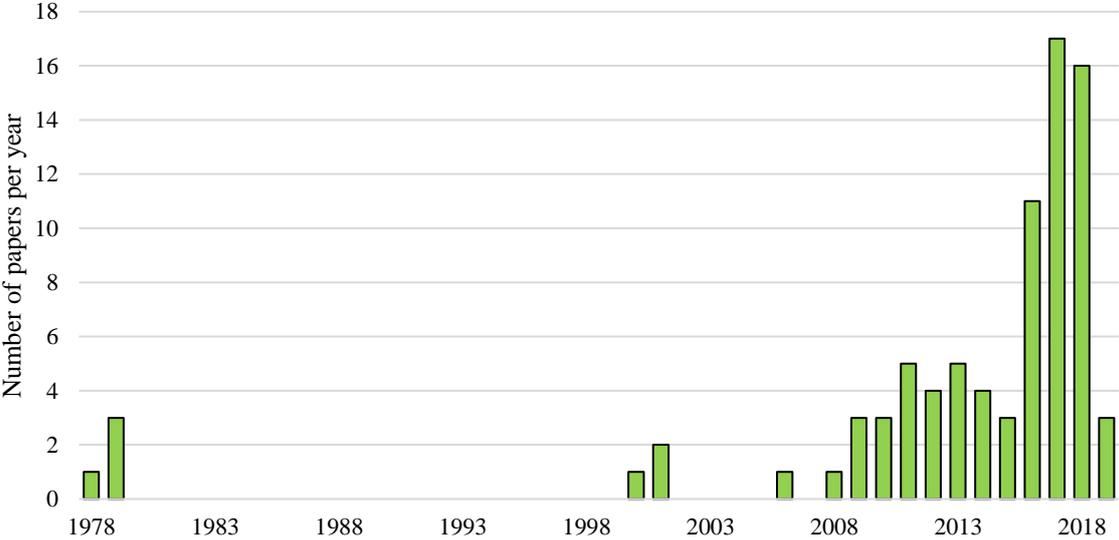

*Figure 3. Number of papers per year in area of renewable energy sources complementarity considered in this review*

Research case studies on the complementarity concept have not been spatially uniformly distributed across the world. According to the conducted literature review, the majority of research papers concentrates on Brazil, China, USA and Europe. Such concentration of research is not a surprise considering the fact that the mentioned regions either historically had a large share of renewable generation (e.g.: Brazil with hydropower) or are currently putting a lot of effort into increasing the share of renewable resources in their power systems (Figure 4 and Figure 5). The spatial coverage and statistics presented in this review are the ones found within the search algorithm described in the introduction section, however, it is probable that further works might be available as theses, reports, etc., written in languages different from English.

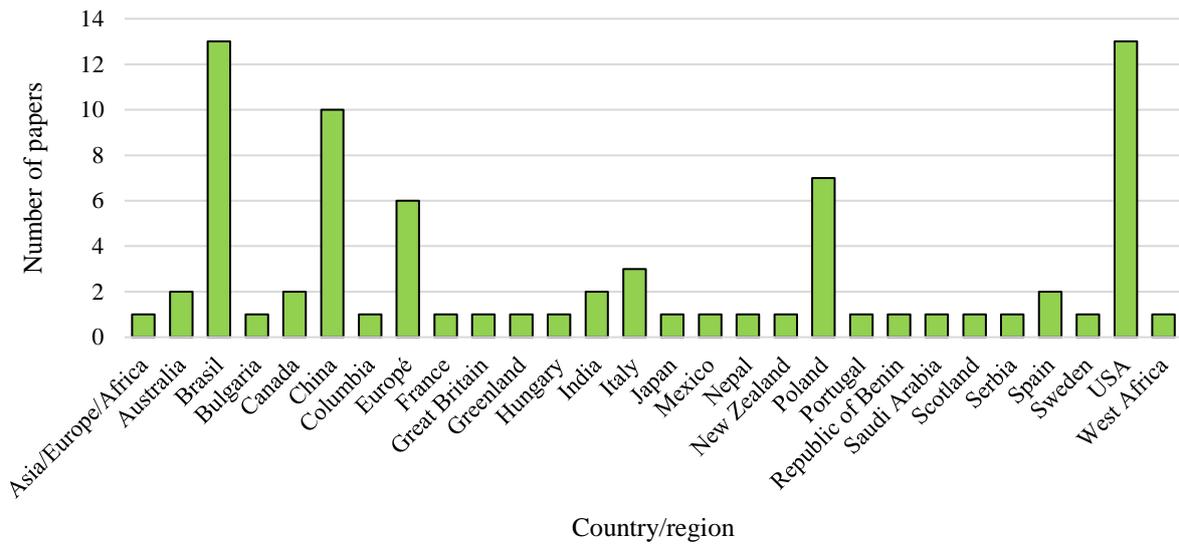

*Figure 4. Number of papers per country/region which investigated the concepts related to energetic complementarity of VRES*

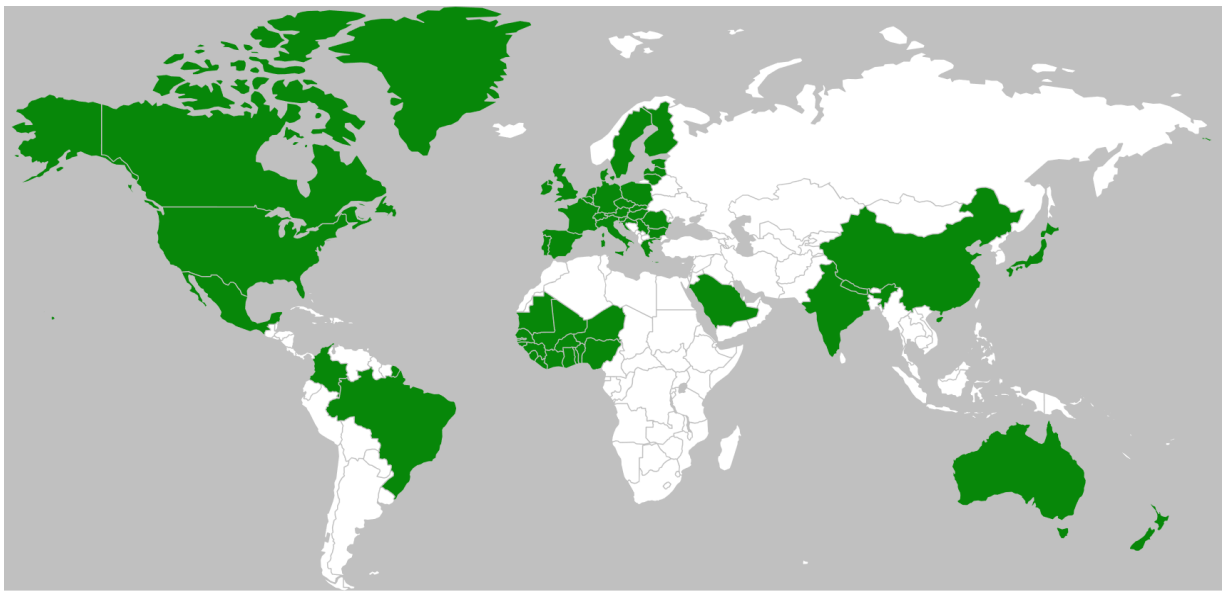

*Figure 5. Spatial coverage of studies considered in this review on renewable resources complementarity*

## 4. Structure of research on complementarity

In this section, we briefly present the reader with the overall statistics regarding the documental database created for this review on renewables complementarity. Overall, the research is dominated by analyses focusing on complementarity between two selected renewable energy sources (over 60% of papers), as seen in Figure 6. Out of these, 34 papers focused on solar-wind complementarity, whereas the remaining works evaluated complementarity between solar-hydro and wind-hydro resources. Research on complementarity between more than two renewable sources is gaining popularity in recent years, however, most of these studies focus on complementarity in terms of optimal sizing and/or operation of solar-wind-hydro systems.

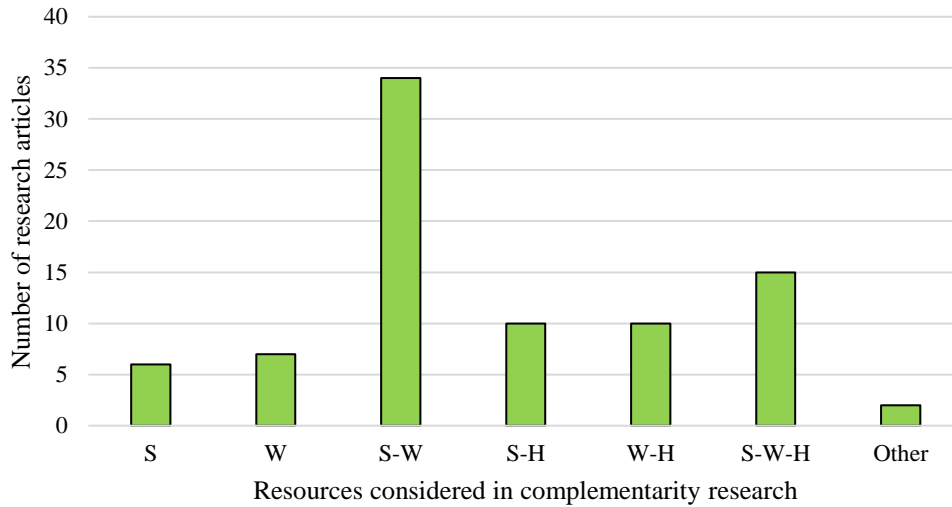

*Figure 6. Number of studies investigating different complementarities. Other refers to combination of solar and/or wind with biomass, wave. – for more details please see Section 6*

One crucial component of any energy system analysis is the temporal resolution to analyze the balance between power supply and demand. Most models for power system operation analysis are based on time series with hourly or sub hourly (15 minutes) time steps. From the documents analyzed in this paper, it is possible to observe that the majority of those works made their analysis based on time series divided in hourly time steps (See Figure 7). It is important to mention that a significant part of the papers included in this review on complementarity investigated this concept for multiple time scales.

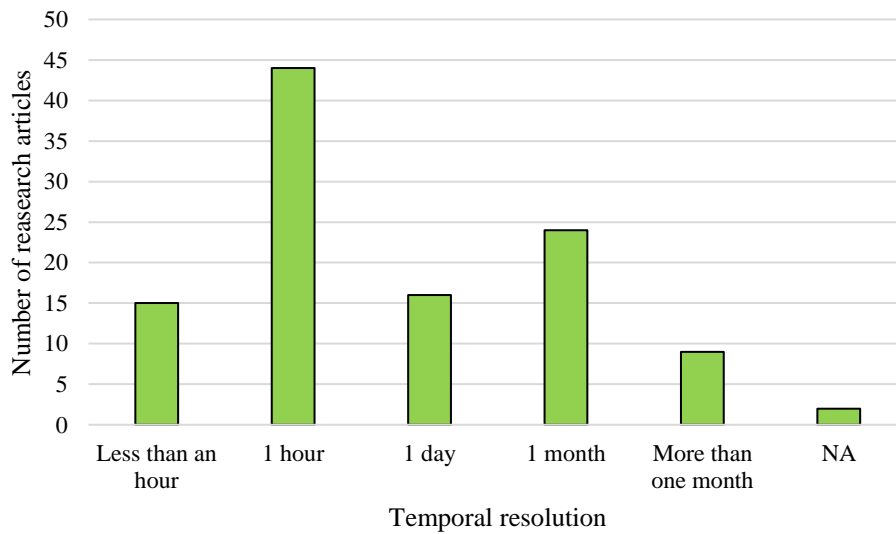

*Figure 7. Temporal resolution of complementarity analysis. Please note that multiple studies considered more than one time-scale. More than one month –refers to inter-annual intra-annual variability.*

Time series for assessing availability and variability of VRES can be obtained from various sources. The most commonly used are ground measurements, satellite measurements and numerical models/reanalysis.

From the consulted references it was found that more than half of the considered studies were based on time series created from ground measurements, whereas the remaining papers relied on mixed data sources or single satellite or reanalysis data sources. A reanalysis is a data assimilation project that combines a physical model with historical observational data into a single consistent dataset. Within the consulted literature, studies that relied on reanalysis were mostly based on Numerical Weather Prediction (NWP) and physical models, *i.e.* numerical models that use physical laws to simulate the atmosphere, such as MERRA-2.

## 5. Quantifying energetic complementarity: indices, metrics and other approaches

Since the early works about energetic complementarity between VRES, authors have been trying to assess this complementarity by means of statistical metrics and other indices. This assessment has become more relevant with the current trend of increasing renewables penetration in national power grids, without affecting reliability and optimizing the financial resources available.

One of the first examples of using metrics for evaluating energetic complementarity can be found in the paper by Takle & Shaw (1979) [14]. Besides using superposition to analyze combined solar and wind energy per unit area, these authors evaluated the product of deviations of the daily total from the expected amount for solar and wind resources, using the monthly and annual averages of these results for drawing their conclusions and suggesting some applications and considerations based on complementarity between the two energy sources.

Since then, several works have been conducted on metrics and indices to evaluate complementarity, as evidenced on Table 3. In this section, we will present the most common and relevant metrics, indices and approaches that have been applied in assessing complementarity between renewable energy sources.

### 5.1. Correlation

Correlation is the most widely-used measure of dependence between two randomly distributed variables, and in a broad sense, it can be defined as a metric that directly quantifies how these variables are linearly related [16]. Correlation has been the metric most commonly used in papers dealing with complementarity measurements, and five types stand out: Pearson correlation coefficient, Kendall correlation coefficient, Spearman's Rank Correlation Coefficient, Canonical Correlation Analysis and Cross-correlation.

#### 5.1.1. Pearson correlation coefficient (simple correlation – $r_{xy}$)

Also called the simple correlation coefficient, this coefficient measures the association strength between two variables, with values ranging from -1 to +1. A value of 0 implies that no association exists between the two variables; a positive value indicates that as the value of one of the variable increases or decreases, the value of the other variable has a similar behavior; on the other hand, a negative value signifies that as the value of one variable increases, the value of the other variable decreases, and vice versa [17]. Another definition of the Pearson correlation coefficient is the covariance of the two variables divided by the product of their standard deviations.

When applied to samples, the Pearson correlation coefficient is usually represented as $r_{xy}$. Given a sample of paired data $\{(x_1,y_1),…,(x_n,y_n)\}$, $r_{xy}$ can be calculated as:

$$r_{xy} = \frac{\sum_{i=1}^{n}(x_i - \bar{x})(y_i - \bar{y})}{\sqrt{\sum_{i=1}^{n}(x_i - \bar{x})^2}\sqrt{\sum_{i=1}^{n}(y_i - \bar{y})^2}} \qquad (1)$$

where *n* is the size of the sample, $x_i$ and $y_i$ are the individual sample points of each variable, $\bar{x}$ and $\bar{y}$ are the sample means for each variable.

For energetic complementarity, the best possible value of $r_{xy}$ = -1, that would suggest full complementarity between sources or regions, as it can be easily observed from the lower left chart in Figure 1 and the information in Figure 2. Similarly, the worst possible value in terms of energetic complementarity would be $r_{xy} = +1$, meaning that the sources or regions present the same pattern, or are concurrent, in terms of maximum and minimum values (see upper left chart in Figure 1).

The most common purposes for calculating the Pearson correlation coefficient regarding energetic complementarity has been: 1) conducting an statistical analysis for evaluating if the renewable energies available in one region could allow the configuration of efficient power systems based on renewables ([15], [17], [18], [19], [20], [21], [22], [23], [24], [25], [26], [27], [28], [29], [30], [31], [32], [33], [34]); 2) as a tool for improving the operation or planning of existing power plants or systems ([12], [13], [35], [36], [37], [38]); 3) as part of the set of equations, parameters and inequalities in an optimization model ([39], [40], [41], [42], [43], [44], [45], [46]).

### 5.1.2. Kendall correlation coefficient (Kendall's Tau – $\tau(x, y)$)

Another type of correlation coefficient, the Kendall correlation coefficient, usually called Kendall's Tau, with notation $\tau(x, y)$, is a non-parametric measure of the rank dependence between two sets of random variables [16]. The Kendall correlation coefficient can be estimated by:

$$\tau(x, y) = \frac{(C - D)}{n(n - 1)/2} \qquad (2)$$

In equation (2), each sample size is $n$, and C is the total number of concordant pairs (ranked vs. real) for both sets of data, and D is the number of discordant pairs. The denominator is the total possible number of paired combinations, so the coefficient must be in the range $-1 \leq \tau(x, y) \leq 1$. Therefore, in terms of energetic complementarity, a perfect agreement between two rankings would yield a $\tau(x, y) = 1$, meaning a concurrent behavior between resources. Consequently, if one ranking is the reverse of the other, then $\tau(x, y) = -1$, indicating the best possible complementarity between the sources.

There are some authors that have applied the Kendall correlation coefficient for energetic complementarity assessments. Denault et al. (2009) [38] used it as one of the copulas for modeling the dependence between wind and hydropower resources in the province of Quebec, for evaluating the possible effect of wind power in reducing the risk of water inflow shortages. Xu et al. [47], in their paper assessing the spatial and temporal characteristics of wind and solar complementarity in China, employed the Kendall rank correlation coefficient as the dependence measure and regionalization index. A recent paper by Han et al. [48] compared the results obtained by the method proposed by them with Kendall's tau, for describing the complementarity between three renewable sources, including fluctuation and ramp effects in their calculations.

### 5.1.3. Spearman's rank correlation coefficient (Spearman's rho – $\rho_S(x, y)$)

This correlation coefficient is another measure of rank dependence. For a sample, the Spearman's correlation coefficient can be described as simply the Pearson correlation applied to ranks [49]. For a distribution or an infinite population, it is required to transform both variables by their univariate marginal cumulative distribution functions (CDF), allowing to compute $r_{xy}$ for the transformed variables [50]. This transformation of a random variable by its CDF is similar to computing the ranks of a variable in a finite sample. In other words, Spearman's coefficient removes the relative size of the two variables, and the dependence is measured between the transformed variables [16]. As with the other types of correlation, $\rho_S(x, y)$ values will always be within the interval $-1 \leq \rho_S(x, y) \leq 1$.

For a sample of size n, the calculation of $\rho_S(x, y)$ requires that each $x_i$ and $y_i$ value have to be converted to ranks *rank $x_i$,* and *rank $y_i$*:

$$\rho_S(x, y) = \frac{covariance(rank\ x\ ,\ rank\ y)}{\sigma_{rank\ x}\ \cdot\ \sigma_{rank\ y}} \tag{3}$$

where $\sigma_{rank\ x}$ and $\sigma_{rank\ y}$ are the standard deviations of the corresponding ranks.

For studies related to energetic complementarity, the Spearman's Rank Correlation Coefficient has been used by Denault et al. (2009) [38] along with $r_{xy}$ and $\tau\ (x,\ y)$ as the copulas for assessing dependence between wind and hydropower resources in Quebec. Using Pearson's and Spearman's correlation coefficients, Cantão et al. (2017) [36] produced wind and hydropower complementarity maps for the entire Brazilian territory based on the weather stations used to create Voronoi cells (or Thiessen polygons).

### 5.1.4. Canonical correlation analysis (CCA)

A technique developed in the 1930's, the canonical correlation analysis (CCA) can be described as a multivariate statistical technique used for identifying possible links between sets of multiple dependent variables and multiple independent variables [51]. Some examples are the relation between governmental policies and economic growth indicators, relation of price variables (initial price, salvage value) of a car and its features, relation between job performance and company characteristics. The CCA method is fully described in [52].

With the southern half of the Iberian Peninsula as case study, Santos-Alamillos et al. (2012) [53] used CCA with the aim of finding the optimal distribution of wind and solar farms over the region, while keeping a regular energy input into the power system, using coupled spatiotemporal canonical patterns for their analysis. In a follow-up paper, Santos-Alamillos et al. (2015) [51], this time using the region of Andalucia as a case study, used Principal Component Analysis (PCA) coupled with CCA to evaluate if a combination of wind power and concentrating solar power (CSP) could provide an adequate baseload power to the region.

### 5.1.5. Cross-correlation

The cross-correlation function is a measure of similarity that compares two component series of a stationary multivariate time series, where a delayed similarity exists [50] [20]. For two real valued sequences $x$ and $y$ of the same time series and delayed by $h$ time units, cross-correlation values range from -1 to 1 and can be found by means of the expression:

$$\rho_{x,y}(h) = \frac{\sum_t [x(t) - \bar{x}] * [(y(t-h) - \bar{y}]}{\sqrt{\sum_t [x(t) - \bar{x}]^2}\ \sqrt{\sum_t [y(t-h) - \bar{y}]^2}} \tag{4}$$

Cross-correlations can help understanding the relation between the component series or how they are influenced by a common factor. However, like all correlations, they are only a statistical measure of association, not causation, therefore, determining causal relationships require further knowledge and analysis [50].

The report by Justus (1979) [15] summarized the results from a series of studies made in 1970's of wind and power distributions for large arrays of wind turbines in United States. In those works, cross-correlation was the main metric employed for assessing spatial complementarity between pair of sites.

In studies related to assessing energetic complementarity, cross correlation has been used in studies for evaluating the possible benefits of distributed wind power generation in Europe [5], measuring the complementarity between load demands and wind and solar resources in Australia [30], PV power fluctuations in the Iberian peninsula [54], and calculating complementarity between renewable energy resources in Brazil [55], [56].

5.2. Indices

An index is a metric used to summarize a set of features in a single value. One index conceived specifically to assess energetic complementarity was proposed by Beluco et al. (2008) and it was tested for analyzing energetic complementarity between solar and hydropower resources in the state of Rio Grande do Sul, Brazil [57]. The time-complementarity index created by the aforementioned authors could be calculated as the product of three partial indices: 1) Partial time-complementarity index (which evaluates the time interval between the minimum values of two sources); 2) Partial energy-complementarity index (which evaluates the relation between the average values of two sources); 3) Partial amplitude-complementarity index (which assesses the differences between maximum and minimum values of the two energy sources). Each one of the partial indices ranges between 0 and 1, therefore, a value of 0 for the time complementarity index indicates that both resources are concurrent, and a value of 1 suggests full complementarity. This index was used in other papers as the energetic complementarity metric, mainly for assessing or reducing energy storage requirements, for the spatial representation of complementarity or for evaluating complementarity in other regions [58], [59], [60], [61], [62], [63], [64], [65], [66].

Li et al. (2011) [67] using the state of Oklahoma as case study, proposed the Complementarity Index of Wind and Solar Radiation (CIWS), an index that incorporates geographic regression modeling and principal component analysis (PCA) for investigating the impacts of diverse geographic factors on complementarity.

In their scheduling optimization model, Zhu et al. (2018) [68] combined wind, solar and hydro power output, and defined this ensemble as a virtual power (VP), according to their complementary features. Then, the capacity of the VP output to follow the load is measured by the load tracking index. Lower values of this index indicate a better performance by the VP, thus, the minimization of the load tracking index is the objective function of this model for assessing energetic complementarity in multiple time-scales.

5.3. Metrics related to failures and reliability

In terms of power supply, a failure can be defined as any situation where the total power supplied by the system composed by the generating units and energy storage devices is less than the load demand. Some authors have assessed energetic complementarity from that perspective. Stoyanov et al. (2010) [69] quantified the number of times and total hours of load faults for a case study in Bulgaria assessing if complementarity between solar and wind resources followed the electrical consumption demands.

Beluco et al. (2012) [58] used a failure index to evaluate the performance of a PV-hydro hybrid system, and from their results they concluded that a smaller failure index is associated to a higher temporal complementarity between the resources.

Assessing the potential of energetic complementarity for increasing system's reliability was one of the first research interests on this subject [13]. Since the early works by Kahn [12], [13], a common metric used in papers assessing the reliability of hybrid power systems is the loss of load probability (LOLP), which can be defined as the probability for a system of being unable to meet the load demand in a given time. Indirectly, this metric accounts for one of the main concerns about renewables and their complementarity: the power output fluctuation. An expression for calculating LOLP in terms of percentage is given by:

$$LOLP = \frac{\sum_{t=1}^{T} E_t^{Def}}{\sum_{t=1}^{T} E_i^{D}} * 100\% \qquad (5)$$

where $E_t^{Def}$ is the energy deficit and $E_i^{D}$ is the energy demand for time intervals from t = 1 to t = T. This metric was used by Schmidt et al. (2016) [70] as a constraint in the optimization model. The LOLP parameter was also employed in the paper by Jurasz et al. (2018) [37] in their analysis on how complementarity affects the power system reliability.

By using several descriptive statistics and other techniques like correlation and linear regression, Shaner et al. (2018) [33] analyzed how the geophysical variability of solar and wind resources affects the system's reliability that can be achieved by different mixes of these two sources. Their findings indicate that energy storage and electricity transmission infrastructure requirements would be a function of the generation mix.

5.4. Assessments based on fluctuations

One of the main concerns related to increasing the fraction of variable renewables in large scale power grids is the disturbance caused by significant oscillations of these sources in time. Based on this, some authors have evaluated the potential of spatial and temporal energetic complementarity for limiting or avoiding these fluctuations. The paper by Gburčik et al. (2006) [71] used the Serbian territory for assessing how complementary regimes of solar and wind energy could be used for reducing these output fluctuations in national grids.

The paper by Murata et al. (2009) [21] investigated the relationship between the largest output fluctuation by means of output fluctuation coefficients, using solar power information from 52 sites in Japan. Their findings suggest that the largest output fluctuation of distributed photovoltaic generation can be predicted by using these output fluctuation coefficients and geographical correlation. Marcos et al. (2012) [54] demonstrated that short-term power fluctuations produced by a set of large PV plants geographically dispersed are considerably diminished when compared with a single PV of the same capacity of the ensemble; both in terms of the largest output fluctuation and the relative frequency.

The ramp rate is a common metric in power generation that expresses how quickly the power output changes over time, and is usually expressed in MW/min. This parameter is established to keep an adequate balance between power supply and demand, preventing undesirable effects in the power system and grid due to these rapid fluctuations in loading or discharge, and their impact on the system's reserve [72]. The ramp rate offers a simple metric for analyzing power transients [73], and because of this, some authors have included ramp rates assessments in their energetic complementarity studies. Some examples can be found in [23], [72] and [73].

The smoothing effect refers to the reduced variability if several locations of power generation are aggregated. It is a common terminology in investigations on variability and commonly studied in a variety of papers. Liu et al. [24] assessed this feature by investigating the duration curves of the single sites and the duration curve of their sum. They found a considerably smaller spread of values, particularly for wind. Hoicka and Rowlands [74] quantified smoothing of the joint generation from wind and solar resources compared to single sources in Ontario, Canada and found that instances of high and low production are less frequent when the complementary behavior is considered.

Krutova et al. [75] evaluated the smoothing effect of Wind-PV mixes in an optimization model, using as case study the entire Afro-Eurasia region. They found a significant balancing potential on these continent-wide scales. Jurasz and Ciapala [76] observed that is possible to smooth the energy demand curve by means of complementarity, as in their paper they evaluate a hybrid PV-run-off-river system. Spatial and temporal wind and solar power characteristics are analyzed by Tarroja et al. [73], [77]. Their findings indicate that size and spatial distribution of these power plants significantly reduces the magnitudes of hourly power fluctuations.

In the paper by Berger et al. [78], they are able to show that low wind power production events can be counterbalanced on a regional scale by taking advantage of the different wind patterns across the region (western Europe and southern Greenland in their case study). Their findings evidenced that wind power production on different continents might decrease the number of low wind power production events, making a case for evaluating the potential benefits of intercontinental electrical interconnections.

### 5.5. Complementarity and optimization models

One of the main research topics related to hybrid power systems based on renewables is finding the optimal mix and operation of these resources (in terms of minimizing capital and/or operational cost, reducing energy storage requirements, etc.). Because of the abundant literature on the subject, an extensive review related to optimal mix of renewables in hybrid power systems falls out of the scope of this paper. Nevertheless, some works that include the explicit or implicit assessment of energetic complementarity in their optimization models are worth mentioning.

Heide et al. (2010) [79] quantified the seasonal optimal mix between wind and solar power for Europe, along with the seasonal storage needs. Nikolakakis & Fthenakis (2011) [80] analyzed the maximum accomplishable penetration of renewable-energy in the power grid of the state of New York, USA. Steinke et al. (2013) [81] prepared a simple model for estimating the energy storage requirements for 100% renewable scenarios in Europe. Similarly, Solomon et al. (2014) [82] examined the storage requirements and maximum possible penetration from intermittent renewables into the electricity grid for the state of California, USA. In a follow-up paper [83], the same authors extended the model, aiming to optimize penetration, minimize energy storage and backup capacity requirements, as well as analyzing dispatch strategies.

Ramos et al. (2013) [84] proposed a model for analyzing hydro-wind complementarity through portfolio optimization and optimal energy allocation strategy, this is, maximizing profit and minimizing risk exposures. Schmidt et al. (2016) [70] assessed how much greenhouse gas emissions could be reduced by optimizing the daily dispatch of renewables in Brazil, along with the loss of load probability. François et al. (2016) [39] presented a paper estimating the optimal mix of renewables for 12 European regions, including run-of-the-river energy in their considerations. The same authors [40] proposed a model for analyzing solar and run-of-the-river complementarity across different temporal scales using as one of the indicators the theoretical storage requirements for balancing generation and load.

The paper by Kies et al. (2016) [85] identified the optimal share of wave power in a 100% renewable power system on the Iberian Peninsula, with this optimal share defined by minimizing backup energy requirements. François et al. (2017) [86] estimated the effect of the hydrological prediction methods on the assessment of the optimal share of solar power and run-of-the-river hydro at different time scales. Ming et al. (2017) [87] proposed a model for the optimal dimensioning of a utility-scale PV plant, considering its joint operation with a hydropower plant and downstream water level. Chattopadhyay et al. (2017) [88] presented a model for determining the backup power and energy storage requirements for a fully renewable European power system, emphasizing on the effect of the PV module configuration. A similar approach of adjusting the PV module configuration for improving complementarity was also used by [41].

The model of Jurasz & Ciapała (2017) [76] was designed to maximize the volume of energy from PV and hydropower used for serving the load, subject to the constraints that both energy deficits and energy surpluses would not exceed 5% of the energy demand. The model proposed by Zhang et al. (2018) [89] aimed to make the most of the complementary behavior of the available VRES, with the objective function of minimizing the excess of wind and solar power, as well as improving energy storage in hydropower reservoirs. Aza-Gnandji et al. (2018) [45] used Particle Swarm Optimization (PSO) for obtaining the optimal geographical coordinates that would minimize the correlation (thus maximizing complementarity) between wind and solar energy in Benin Republic.

More recently, da Luz & Moura (2019) [90] created a model for optimizing the mix between solar, wind, hydro and biomass power, from a complementarity perspective, in order to supply Brazil's monthly and hourly load demand based on projections for the year 2050. Naeem et al. (2019) [46] used the Pearson correlation coefficient for assessing solar-wind complementarity in an optimization model whose objective function aims at maximizing economic benefits of a microgrid, by means of exploiting such spatial and temporal energetic complementarity. Some other models dealing with optimization based on complementarity can be found in [42], [44], [68], [75], [91], [92], [93] and [94].

### 5.6. Other relevant metrics

As seen along this section of the paper, complementarity assessments have been conducted through different approaches. There are some that are not possible to categorize in the previous subsections, but worth mentioning in this document.

Glasbey et al. (2001) [95] created a method for the statistical modelling of spatio-temporal variations of global irradiation on a horizontal plane, using covariance as the main metric for assessing the impact of time lag and distance on irradiation complementarity in two sites.

The Power Spectral Density (PSD), which is a measure of power content versus frequency, has been used by [22], [27], [73] and [77] for characterizing the observed variability of wind and solar power plants as function of different time scales and locations.

Risso & Beluco (2017) [64] proposed a method for performing a graphical representation of temporal complementarity of resources at different locations, by means of a chart of complementarity as a function of distance, using a hexagonal cell network for dividing the case study region. In a follow-up paper [65], the method was extended, with the graphical representation now portrayed as complementary roses, with the length of the petals denoting the distance to another cell and their color the magnitude of energetic complementarity between these cells.

Spatial and temporal complementarity (synergy) of wind and solar resources in Australia was assessed by Prasad et al. (2017) [96]. The Relative Coefficient of Variation was the main metric employed for evaluating the variability of these renewables, and besides this, the method mostly consisted in measuring the occurrence of solar and wind resource above a minimum threshold. Results were presented as a set of maps.

The concept of critical time windows, which represent periods within the time series with low average capacity factors, are proposed by Berger et al. (2018) [78] for the systematic assessment of energetic complementarity over both space and time. These critical time windows provide an accurate description of extreme events within the time series, while retaining chronological information. These authors also propose a criticality indicator that quantifies the fraction of time windows during which generation from variable renewables is below a certain threshold, allowing a comprehensive evaluation of energetic complementarity in the different locations over arbitrary time scales.

The local synergy coefficient was a metric employed by Zhang et al. (2018) [72] for representing the mutual complementarity between VRES at one site, based on the normalized capacity factors of the sources.

The stability coefficient $C_{stab}$ was a metric developed by Sterl et al. (2018) [97]. It can be defined as a measure that quantifies the added value of one VRES to balance the daily power output from another VRES. In their paper, these authors assess the capacity of wind power for balancing PV power in West

Africa, based on diurnal timescales of the capacity factors of a hybrid power system with equal installed capacity of PV and wind power.

### 5.7. Assessing complementarity between more than two sources

From the previous paragraphs, it can be observed that complementarity is usually measured between two VRES. However, there are authors that have extended the existing methods in literature, in order to allow the assessments of energetic complementarity between more than two sources. Borba and Brito [63], extending on the method presented in [57], proposed a dimensionless index for calculating temporal complementarity between two or more energy resources. In their paper, the complementarity index is calculated as the ratio between the actual generation discarding excess power, and the average generation.

The complementarity between wind, solar and hydropower generation is evaluated by Han et al. [48] by means of comparing fluctuations and ramp ratios between individual and combined power generation. The method was tested using a region in China as case study, and their findings suggest that complementarity can be improved by adjusting the proportion of solar and wind power.

## 6. Tabular summary of relevant papers on the concept of complementarity

This section presents table, which summarizes the most relevant information regarding energetic complementarity according to the consulted literature.

*Table 3. Overview of complementarity studies.*
*Resources: S – solar, W – wind, H – hydro.*
*Time-scale: h – hourly, d- daily, m – monthly, i – interannual.*
*Data sources: M – measurements, S – satellite, R – reanalysis.*
*Comp. – Type of complementarity: S – spatial, T – temporal, S/T – spatio/temporal.*
*NA – not available/specified.*

| Resources | Time-scale | Data sources | Comp. | Complementarity metric/approach | Region | Findings/highlights/comments | Ref. |
|---|---|---|---|---|---|---|---|
| W | h | M | S/T | ●*Pearson's*<br>●*LOLP* | USA | ● Investigation of the wind potential to replace conventional generators<br>● Fundamental work on spatial smoothing increases overall reliability of wind generation | Kahn, 1978 [12] |
| W | NA | M | S/T | ●*Pearson's*<br>●*Cross correlation* | USA | ● Spatial cross-correlations for climatological reduction | Justus and Mikhail, 1979 [15] |
| S-W | d | M | T | ●*Product of deviations from expected energy*<br>●*Energy density superposition* | Central Iowa (USA) | ● Daily complementarity is insignificant, whilst seasonal enables much greater compensation | Takle and Schaw, 1979 [14] |
| W | h | M | S/T | ●*Pearson's* | California (USA) | ● Reliability of variable generators results from geographical dispersal, and it is limited by regional saturation and wind speed correlations | Kahn, 1979 [13] |
| S-W | h, d, m | M | T | ●*Pearson's* | Dhahran (Saudi Arabia) | ● Correlation indicates efficient use of hybrid systems.<br>● Input data with exceptional 1-minute resolution. | Sahin, 2000 [18] |
| W | h | M | S/T | ●*Pearson's Cross Correlation* | Europe | ● Smoothed wind generation can cover 20% of European demand, whilst curtailing only 10% of the generation. | Giebel, 2001 [19] |
| S | 1/6h, 1/1200h | M | S/T | ●*Covariance between sites* | Pentland Hills and Edinburgh (Scotland) | ● Study shows how knowledge about the operation of dispersed solar generators can be used to control the power system | Glasbey et al., 2001 [95] |

| Resources | Time-scale | Data sources | Comp. | Complementarity metric/approach | Region | Findings/highlights/comments | Ref. |
|---|---|---|---|---|---|---|---|
| S-W | m | M | T | ● *Output Fluctuation* | Serbia | ● The fluctuation output of solar-wind generators can be smoothed by their cumulative and complementary operation | Gburčik et al., 2006 [71] |
| S-H | m | M | T | ● *Beluco index* | Rio Grande de Sul (Brasil) | ● Introduction of a comprehensive index for complementarity assessment including the complementarity in time, power and energy | Beluco et al., 2008 [57] |
| S-W | ½ h | M | T | ● *Pearson's Cross Correlation* | Australia | ● This study shows that combined solar/wind generation can reliably cover load when grid faces peak demand | Li et al., 2009 [20] |
| S | 1/60h | M | S | ● *Pearson's* ● *Output Fluctuation* | Japan (52 sites) | ● By introducing the output fluctuation index the paper proposes a method to estimate the fluctuations of dispersed generators | Murata et al., 2009 [21] |
| W-H | h | M | S/T | ● *Pearson's Kendall's* ● *Spearman* | Canada | ● The addition of wind capacity to the hydropower dominated system minimizes the deficit of inflow deficit | Denault et al., 2009 [38] |
| W-S | h | M | T | ● *Quantity and duration of faults* | Bulgaria | ● One of the first approaches for general methodology of power systems coupled with larger scale renewable generators | Stoyanov et al., 2010 [69] |
| W | 1/60h and 1/4h | M | S/T | ● *Pearson's* ● *Power Spectral Density* | USA | ● The results indicate that simple interconnection of power plants does not ensure reliability of supply<br>● 3-10% of installed capacity can be provided as firm power due to the dispersion and interconnection effect | Katzenstein et al., 2010 [22] |
| S-W | h | R | S/T | ● *Optimal mix* | Europe | ● Study presents various scenarios of solar and wind in autonomous mode, connected to PSH or hydrogen facility, connected to fossil and nuclear power stations | Heide et al., 2010 [79] |
| S-W | h | M | T, S/T | ● *Smoothing effect* | Ontario (Canada) | ● Spatial distribution and hybridization in single location, between locations smoothens the joint power generation profile<br>● No benefits/drawbacks are visible when comparing solar and wind in different | Hoicka and Rowlands, 2011 [74] |

| Resources | Time-scale | Data sources | Comp. | Complementarity metric/approach | Region | Findings/highlights/comments | Ref. |
|---|---|---|---|---|---|---|---|
| | | | | | | locations with both of them in a single location | |
| S-W | m | M | T, S | ●*CIWS* | Oklahoma (USA) | ● New interesting index is proposed which is suitable for both GIS and hybrid optimization related research. | Li et al., 2011 [67] |
| S-W | h | M/R | T | ●*Optimal mix* | New York (USA) | ● This paper does not directly investigate the complementarity concept but it is clearly used in the modelling procedure | Nikolakakis and Fthenkais, 2011 [80] |
| W | 1/6h | R | T, S | ●*Smoothing effect* ●*Power Spectral Density* | California (USA) – one site | ● The relative wind power fluctuations can be reduced by dispersing the wind generators over geographic | Tarroja et al., 2011 [77] |
| S-W | h | M/R | S/T | ●*Power ramp* ●*Pearson's* | Sweden | ● Wind and solar are negatively correlated on all time scales from hours to years however the strongest complementarity is observed for monthly sums | Widen, 2011 [23] |
| S-W | h, m, i | R | T, S, S/T | ●*Canonical correlation analysis* | Spain | ● Optimal location of solar and wind parks will significantly reduce the power output fluctuations ● Balancing between solar and wind exists but exhibits a significant seasonality | Santos-Alamillos et al., 2012 [53] |
| S-H | N/A | N/A | T | ●*Beluco* ●*Failure index* | N/A | ● The complementarity characteristic can be used for designing the hydro-PV hybrid stations ● A supply reliability is presented in the function of complementarity index | Beluco et al., 2012 [58] |
| S | 1/3600h | M | S | ●*Cross Correlation* ●*Output Fluctuation* | Spain (South of Navarra) | ● Large output fluctuations and fluctuations frequency is reduced by employing spatial dispersion of PV plants ● 6 kilometers distance between plants is found to be sufficient to ensure uncorrelated power fluctuations | Marcos et al., 2012 [54] |
| W-H | m | M/N | S/T | ●*Optimal mix* | Brazil – multiple locations | ● Paper proposes a model to analyze hydro-wind complementarity and the portfolio effect on financial profits and risk exposures | Ramos et al., 2013 [84] |
| S-H | m | M | T | ●*Beluco index* | N/A | ● Papers provides a theoretically analysis of the effects of complementarity which | Beluco et al., 2013 [59] |

| Resources | Time-scale | Data sources | Comp. | Complementarity metric/approach | Region | Findings/highlights/comments | Ref. |
|---|---|---|---|---|---|---|---|
| | | | | | | are: 1) reduction of failing to meet load demand, 2) reduction of storage capacity | |
| S-W | h | M | S/T | ● *Pearson's*<br>● *smoothing effect* | China – 22 locations | ● Paper investigates complementarity between wind and solar PV as compared to single sources alone | Liu et al., 2013<br>[24] |
| S-W | Several time scales | R | S/T | ● *Optimal backup and energy storage capacity* | Europe – 16 locations | ● The authors study the interplay between grid and storage extensions for integrating intermittent renewable energies in Europe using a simple model. | Steinke et al., 2013<br>[81] |
| S | 1/12h | M | S | ● *Power Spectral Density*<br>● *Ramp rate* | USA – 8 locations | ● In the paper, a spectral method is used to distinguish between cloud- and diurnal-cycle induced transients. | Tarroja et al., 2013<br>[73] |
| W-S | h, d, m | M | T, S, S/T | ● *Pearson's* | Italy | ● Complementarity between wind and solar resources for a sample year in Southern Italy is assessed. Authors find strong advantage of complementarity. | Monforti et al., 2014<br>[26] |
| S-W | h | R/S | S/T | ● *Optimal mix and energy storage capacity* | USA – multiple locations | ● Paper estimates storage size, backup requirements and transmission requirements and their interplay to design a low-cost renewable system. | Solomon et al., 2014<br>[82] |
| W-H | h | M/R | S/T | ● *Beluco index* | Brazil – multiple locations | ● Complementarity between wind and hydro resources in Brazil is assessed. Paper provides a map of spatially resolved complementarity in time. | Eifler Neto et al., 2014<br>[60] |
| W-S-H | m | M | T | ● *Qualitative analysis* | Nepal | ● Complementarity of wind, PV and hydro is investigated. Results indicate that a well-planned transmission system could exploit the complementarity. | Kunwar, 2014<br>[98] |
| W-S-H | m, i | R | T | ● *Pearson's* | Colombia | ● MERRA reanalysis data is used to study the complementarity of wind/solar and hydro resources in Colombia | Ramírez, 2015<br>[35] |
| S-W (CSP) | h, m, i | R | T, S, S/T | ● *Canonical correlation analysis* | Spain | ● PCA and CCA are used to find optimal locations of CSP plants and wind farms in Southern Spain. | Santos-Alamillos et al., 2015<br>[51] |
| S | 1/60h and 1/30h | M | S | ● *Pearson's* | Gujarat (India) | ● 13 months of observed power production from utility-scale plants in Gujarat, India are used to analyze the | Klima and Apt, 2015<br>[27] |

| Resources | Time-scale | Data sources | Comp. | Complementarity metric/approach | Region | Findings/highlights/comments | Ref. |
|---|---|---|---|---|---|---|---|
| | | | | ●*Power Spectral Density* | | potential of geographic smoothing of solar PV. | |
| W-S | d | M | T | ●*Cross Correlation* | Fernando de Noronha (Brazil) | ● Correlations of wind speed and solar irradiation are analyzed. Results indicate existence of complementarity between persistence properties of both sources. | dos Anjos, et al., 2015 [55] |
| H-W (off-shore) | m | M/S/R | S, T | ●*Pearson's* | Brazil | ● The possibility of complementing hydro power with offshore wind is investigated for Brazil. | Silva et al., 2016 [56] |
| W-H | h | M/S | S/T | ●*Optimal mix, backup and energy storage capacity* | California | ● Demonstrates advantages of wind and solar complementarity for highly renewable grids in combination with energy storage. | Solomon et al., 2016 [83] |
| S-W-H (hydrokinetic) | h, d, m, i | M | T | ●*Pearson's* | Poland – two cities | ● This paper investigates complementarity between solar and hydro-kinetic energy for two sites in Poland. | Jurasz et al., 2016a [29] |
| S-W-H (run-of-river) | d | M/R | T | ●*Pearson's* ●*Optimal mix* | Europe – 12 regions | ● The paper studies the Integration of run-of-the river (RoR) power in the solar/wind power mix and shows that this integration significantly increases the penetration rate. | François et al., 2016a [39] |
| S-H (run-of-river) | h, d, m | M | T | ●*Pearson's* ●*Optimal energy storage capacity* | North-Eastern Italy | ● Run-of-the-river power and solar PV are used to smoothen energy balance in North-Eastern Italy. Results indicate that the optimal share depends on the considered time scale. | François et al., 2016b [40] |
| S-H | d | M | T | ●*Beluco index* | Brazil | ● 50 years of long-term data are used to study seasonal and inter-annual variability as well as solar-hydro complementarity in Brazil. | Mouriño et al., 2016 [61] |
| S-W-H-wave | h | M/R | S/T | ●*Optimal backup* | Spain/Portugal | ● Wave, wind and solar PV resources on the Iberian Peninsula are investigated and the optimal mix with respect to the need for balancing energy is determined. | Kies et al., 2016 [85] |
| S-H | m | M/R/S | T | ●*Pearson's* | Hungary – one site | ● A methodology is developed to assess the complementarity between solar PV | Kougias et al., 2016 [41] |

| Resources | Time-scale | Data sources | Comp. | Complementarity metric/approach | Region | Findings/highlights/comments | Ref. |
|---|---|---|---|---|---|---|---|
| | | | | ●*Optimal configuration* | | and small hydro. In addition, an optimization algorithm maximizes complementarity while allowing for small compromises in solar energy output. | |
| S-W-H | D | M/R/S | S/T | ●*Optimal mix* ●*LOLP* | Brazil | ● Solar and wind sources can reduce the Brazilian power system probability of failing to supply the energy due to large share of hydropower | Schmidt et al., 2016 [70] |
| S-W-H | d, m, i | M/R | T | ●*Pearson's* | Poland – one site | ● Authors study multiannual, monthly and daily of solar, wind and hydro resources in a region in Poland. | Jurasz and Piasecki, 2016 [30] |
| S-W | m, quarterly | R | T, S/T | ●*Pearson's* | Great Britain | ● Findings are bimodal wind-irradiance distribution, weak anti-correlation of wind speed and cloudiness and that more solar than wind power capacity increases variability in summer. | Bett and Thornton, 2016 [28] |
| H-S | h, i | M | T | ●*Optimal mix* | Northern Italy (four catchments) | ● The skill of different hydrological prediction methods to predict complementarity between run-of-the-river hydropower and solar power in mountain basins of the Eastern Italian Alps is studied. Results indicate that performance depends on the temporal scale. | François et al., 2017 [86] |
| S-W | h | R/M | T | ●*Pearson's* | Europe | ● Authors use three years on 100m wind taken from ECMWF analyses/forecast to assess the possibility of combined solar/wind energy use over Europe. | Miglietta et al., 2017 [31] |
| S-H | h | M | T | ●*Optimal mix and operation* | China – one location | ● A cost-benefit analysis is performed to optimise the size of utility-scale PV power. Variation in downstream water level consequently used to restrict PV integration. | Ming et al., 2017 [87] |
| H-W-S | annual | M/S | S/T | ●*Beluco index* | Brazil | ● The complementarity between wind, solar and hydro in the Brazilian state of Rio Grande do Sul is studied. Mathematical dimensionless ratios which focus on intra-annual periods, are proposed to measure complementarity. | Bagatini et al., 2017 [62] |

| Resources | Time-scale | Data sources | Comp. | Complementarity metric/approach | Region | Findings/highlights/comments | Ref. |
|---|---|---|---|---|---|---|---|
| S-W | h | R | S/T | • *Coefficient of variation*<br>• *% of time some thresholds are exceeded* | Australia | • A strong temporal synergy of solar and wind resources in Australia is shown and that greater synergy characteristics are in close proximity to established transmission infrastructure. | Prasad et al., 2017 [96] |
| W-H | m | M | T, S, S/T | • *Pearson's*<br>• *Spearman* | Brazil | • Voronoi diagram (Thiessen polygons) for representing complementarity through correlation maps. | Cantão et al., 2017 [36] |
| S | h, d, weekly | S/R | S/T | • *Optimal mix, backup and energy storage capacity* | Europe | • Paper assess the impact of different PV module configurations on backup and energy storage requirements. | Chattopadhyay et al., 2017 [88] |
| S-H | h | M | T | • *Smoothing effect*<br>• *Optimal mix* | Poland – single location | • A MINLP model for a PV-ROR hybrid power system optimization, avoiding large deficits or energy excesses. | Jurasz and Ciapala, 2017 [76] |
| W-S | h | M | T | • *Kendall's* | China | • Complementarity maps of wind and solar resources were created, using Kendall's tau as the regionalization indicator. | Xu et al., 2017 [47] |
| W-S | h | R | S/T | • *Optimal mix* | India – 10 regions | • Their findings show that energetic complementarity, transmission lines and backup can be used to overcome difficulties related to monsoon. | Gulagi et al., 2017 [91] |
| H-W | m | N/A | S/T | • *Beluco index*<br>• *Complementarity charts based on distance* | Brazil – Rio Grande do Sul | • A method is outlined for the spatial representation of temporal complementarity as a function of distance between power plants. | Risso and Beluco, 2017 [64] |
| S-W-H | h, d (hydro) | M | S/T | • *Pearson's*<br>• *Optimal mix* | Brazil – Rio de Janeiro | • An LP model for optimizing the mix of three different renewables.<br>• A correlation matrix was created for the 13 plants considered in the case study. | de Oliveira Costa Souza Rosa et al., 2017 [42] |
| S-W | h | R | S/T | • *Smoothing effect*<br>• *Optimal backup* | Africa/Europe/Asia (selected locations) | • Paper evaluates the balancing potential of a solar and wind power based supergrid covering Eurasia and Africa. | Krutova et al., 2017 [75] |
| S-W-H | h (not clear) | M | T | • *Beluco index extended to three sources* | Rio Grande do Sul (Brazil) | • Based on the index created by Beluco et al. (2008), the method allows the calculation of complementarity between more than two sources. | Borba and Brito, 2017 [63] |

| Resources | Time-scale | Data sources | Comp. | Complementarity metric/approach | Region | Findings/highlights/comments | Ref. |
|---|---|---|---|---|---|---|---|
| S-W | h, m, i | S/M | S | ●*Pearson's* | Lower Silesia (Poland) | ● Results suggests wind and solar complementarity combined with PSH might justify developing a Hybrid power system for the region in study. | Jurasz et al., 2017 [32] |
| S-W | d | S/M | S/T | ●*Pearson's* | Poland – multiple locations | ● Results indicates that complementarity between wind and solar resources in Poland is only significant on a monthly time scale. | Jurasz and Mikulik, 2017 [43] |
| S-W | M | M | S/T | ●*Pearson's* | Mexico | ● Correlation maps were prepared for the entire country and for each season of the year. | Vega-Sánchez et al., 2017 [17] |
| S-W-H | h | M | T | ●*Optimal operation* | Yalong river (China) | ● An optimization model is proposed, aiming at minimizing excess wind and photovoltaic power and maximizing the stored energy. | Zhang et al., 2018a [89] |
| W-S | h | R | S/T | ●*Pearson's* ●*LOLP* ●*Descriptive statistics* | USA | ● Findings indicate that solar and wind power are not enough to provide a highly reliable energy system in continental USA without adequate ancillary infrastructure. | Shaner et al., 2018 [33] |
| W-H | M | M | S/T | ●*Beluco index* ●*Complementarity Roses* | Brazil – Rio Grande do Sul | ● A method is proposed for the representation of spatial complementarity in time in the form of complementary roses. | Risso et al., 2018 [65] |
| H-W-S | h | M | T | ●*Optimal mix and operation* | China | ● The complementary coordinated operation model aims at maximizing the renewables consumption by means of an adaptive simultaneous peak regulation strategy. | Wang et al., 2018 [92] |
| W-S | h | M/R | S/T | ●*Optimal operation* | Lower Silesia (Poland) | ● A model for simulating and optimizing operation of a large scale solar-wind hybrid power system coupled with PSH. ● The local consumption index incorporates grid-related cost. | Jurasz et al., 2018a [93] |
| H-S | h | S | T | ●*Beluco index* | N/A | ● The paper studies the influence of time-complementarity on the energy storage requirements of hydro-PV hybrid power systems. | During Filho et al., 2018 [66] |

| Resources | Time-scale | Data sources | Comp. | Complementarity metric/approach | Region | Findings/highlights/comments | Ref. |
|---|---|---|---|---|---|---|---|
| | | | | | | • HOMER software used for performing simulations. | |
| W | Time windows 1h to 10d | R | S/T | • *Critical time windows*<br>• *Criticality index*<br>• *Smoothing effect* | France – South Greenland | • The critical time windows represent periods of within the time series with low average capacity factors.<br>• The criticality indicator quantifies the fraction of time windows with renewable generation below a defined value. | Berger et al., 2018 [78] |
| S-W-H | h | M | T | • *Pearson's*<br>• *Optimal Operation* | China | • An optimization model is proposed for the complementary operation of a hydro-wind-solar hybrid power system. | Zhu et al., 2018a [44] |
| S-W | h | R/M | T, S | • *Stability coefficient* | West Africa | • The stability coefficient quantifies the complementarity of PV and wind power for balancing power output and reduce energy storage needs. | Sterl et al., 2018 [97] |
| S-W | 1/4h | R/S | T | • *Pearson's*<br>• *LOLP* | 86 locations (Poland) | • Paper assesses how solar and wind resources temporal complementarity impacts on system reliability in terms of covering fixed load.<br>• Evaluation of relations between energy storage, complementarity and system reliability. | Jurasz et al., 2018b [37] |
| W-H | 1/6h (wind), d (hydro) | M | S/T | • *Pearson's* | 2 sites hydro, 15 wind, (New Zealand) | • Paper studies the correlation between the spatiotemporal distribution of renewable energy resources and electricity demand and prices. | Suomalainen et al., 2018 [25] |
| W-S-H | 1/4h, h | M | T | • *Load Tracking Index*<br>• *Optimal scheduling* | China | • Renewable power outputs are bundled as a virtual power (VP).<br>• The Load Tracking Index represents the ability of the VP to efficiently track the load curve. | Zhu et al., 2018b [68] |
| S-W | d | M | S/T | • *Pearson's*<br>• *Optimal locations based on complementarity* | Republic of Benin | • Paper proposes a method for selecting optimal locations for renewable generation based on complementarity, using Particle Swarm Optimization. | Aza-Gnandji et al., 2018 [45] |
| S-W | h | R | S/T | • *Local synergy coefficient* | China – multiple locations | • The paper presents the framework used in the development of an open-source tool, named Quantitative Synergy | Zhang et al., 2018b [72] |

| Resources | Time-scale | Data sources | Comp. | Complementarity metric/approach | Region | Findings/highlights/comments | Ref. |
|---|---|---|---|---|---|---|---|
| | | | | ●*Ramp rates* | | Assessment Toolbox for renewable energy sources. | |
| W-S-H | h | M | T | ●*Optimal mix and operation* | China – single location | ● Propose an optimization model aiming at maximizing renewable generation and steady output of a system based on complementary wind/PV/hydro sources. | Dou et al., 2018 [94] |
| W-S | 1/2h | R/S | T, S | ●*Pearson's* | Texas (USA) multiple locations | ● The findings suggest possible alternatives for making renewables projects able to maximize reliability with minimal investment in storage technologies. | Slusarewicz and Cohan, 2018 [34] |
| W-S-H | h | M | T | ●*Kendall's* ●*Ramp Ratio and Fluctuation rate of combined power generation* | Single location in Northern China | ● Two indices are presented to describe complementarity between two or more energy sources. | Han et al., 2019 [48] |
| S-W-H-biomass | h, m (biomass) | M | S/T | ●*Optimal mix and operation* | Brazil (whole country) | ● The model proposed optimizes the renewable mix and operation of hydropower reservoirs, based on daily and yearly variations. ● Optimal mix is calculated from a complementarity perspective. | da Luz and Moura, 2019 [90] |
| W-S | h | R | S/T | ●*Pearson's* ●*Optimal mix and operation* | Test grid | ● An optimization model uses the complementary characteristics of variable renewables to achieve an economic and reliable operation of a grid-tied microgrid. | Naeem et al., 2019 [46] |

## 7. What is the potential application of the complementarity concept?

The concept of renewable sources complementarity is often used indirectly or without a direct reference to it. Based on the conducted literature review the following applications can be distinguished:

- Hybrid energy sources like solar-wind, solar-hydro which increase the overall system reliability and can reduce the cost of electricity. The underlying principle of hybrid energy sources (utilizing non-dispatchable renewables) is the complementary nature of their energy generation patterns. This phenomenon although crucial is often not explicitly mentioned. As indicated in Jurasz et al. (2018) [37] the varying degree of complementarity can lead to different levels of hybrid system reliability.
- Use of the availability of one source over the course of certain period to reduce the exploitation of another one which is used for different purposes. For example, in hydropower dominated power systems, the reservoirs are often used for energy generation and irrigation purposes (food-energy nexus). In such cases the hydropower generation can be reduced during dry periods by utilizing photovoltaics or wind power (if beneficiary complementary nature of those sources is observed). Another example is the joint operation and scheduling of hydropower-solar/wind stations with biomass facilities. It is a known problem that dry years result in a lower crops harvest which in consequence may cause problems on the supply side of the power stations using biomass. In such situations it may be beneficiary to substitute some hydropower generation by wind/solar and use the water when required for irrigation purposes. The energy will still be generated by the hydropower station although not in such a flexible manner as it used to be.
- Finding the optimal energy sources mix for given power system (tradeoff between curtailment, storage, and cost of energy from different sources – for example even if electricity from PV is more expensive than wind it still might be cheaper to deploy them and use directly during the daylight instead of oversizing wind generation and accompanying electricity storage. Also, the spatial distribution (spatial complementarity) can be considered where wind parks are scattered over larger areas a selection of sites resulting in lower capacity factor may be compensated by lower demand on backup capacity due to the spatial smoothing effect.
- The concept of complementary renewable resources can be used in the energy management strategy for the hybrid systems operating on the electricity markets. For example, solar/wind-hydro stations (with reservoir) can effectively operate on the day-ahead market where bids are required in advance. Bidding the generation of non-dispatchable generators (wind/solar) is purely based on the forecasted generation and is prone to the forecasting errors. A joint operation with complementary hydro station may be feasible option to compensate for the forecasting errors of solar/wind generation.
- As indicated by Kougias [41] and Jurasz & Ciapala [76], the power output from variable generators can be smoothed by their joint operation. For example, an appropriate energy management strategy (for solar-hydro) can smooth the generation patterns and improve the performance of a power system from the perspective of power fluctuations, when compared to generation from only a PV system.

## 8. Conclusions

From the extensive literature review conducted on papers assessing energetic complementarity between renewable sources, the following conclusions and potential research directions can be formulated:

- There are many geographical areas for which VRES energetic complementarity has not been evaluated yet (mostly parts of Africa and Asia);

- Some of the existing complementarity metrics can be extended to consider other aspects related to VRES like the relation between capacity factor and levelized cost of energy, since better complementarity not always equals lower overall cost of the system.
- Future studies should extend the complementarity assessments for allowing the user to understand not only on the statistical relationship (complementarity) between the energy sources, but also to obtain additional information related to the practical application of those metrics;
- Complementarity metrics have been included in several optimization models in order to find the best design and/or operation schedule of hybrid power systems. However, the extent of potential applications can be extended to hydrological models (involving water-energy-food nexus) or power system planning;
- Complementarity metrics should be compared based on the same data sample and their performance should be assessed based on the same criterion to clearly formulate their strong and weak sides;
- Since a majority of complementarity studies focusses on wind/solar/hydro combination, the future research should include some additional renewable sources like wave or tidal energy that have gained recent attraction;
- The research on complementarity should not be based only on the historical datasets, but also consider future climate models and the impact of these changes in renewables complementarity.